\documentclass[prl,twocolumn,floatfix,showpacs ]{revtex4-1}
\UseRawInputEncoding
\usepackage{amsmath}
\usepackage{amssymb}
\usepackage{graphicx}
\usepackage{dcolumn}
\usepackage{bm}
\usepackage[colorlinks=true,linkcolor=blue,anchorcolor=blue, citecolor=cyan,urlcolor=cyan]{hyperref}
\usepackage[mathlines]{lineno}
\usepackage{ulem}
\usepackage{epstopdf}
\begin{document}
\title{Spin ordering-induced fully-compensated ferrimagnetism}
\author{San-Dong Guo$^{1}$}
\email{sandongyuwang@163.com}
\author{Shaobo Chen$^{2}$ and Guangzhao Wang$^{3}$}
\affiliation{$^{1}$School of Electronic Engineering, Xi'an University of Posts and Telecommunications, Xi'an 710121, China}
\affiliation{$^{2}$College of Electronic and Information Engineering, Anshun University, Anshun 561000, People’s Republic of China}
\affiliation{$^{3}$Key Laboratory of Extraordinary Bond Engineering and Advanced Materials Technology of Chongqing, School of Electronic Information Engineering, Yangtze Normal University, Chongqing 408100, China}

\begin{abstract}
Fully-compensated ferrimagnets exhibit zero net magnetic moment yet display non-relativistic global spin splitting, making them highly advantageous for constructing high-performance spintronic devices.  The general strategy is to break the inversion symmetry of conventional antiferromagnets or the rotational/mirror symmetry of altermagnets to achieve fully-compensated ferrimagnets. Here, we propose to induce fully-compensated ferrimagnetism by engineering the spin ordering rather than modifying the lattice structure. Bilayer stacking engineering offers a convenient platform to verify our proposal and readily enables switching between two distinct electronic states by tuning the $\mathrm{N\acute{e}el}$ vector of one layer.  By the first-principles calculations, a bilayer system is constructed with monolayer $\mathrm{Cr_2C_2S_6}$ as the elementary building block to corroborate our proposal. This strategy  can also be extended to inducing altermagnetism via spin ordering  engineering. Our work offers an alternative route to realize non-relativistic spin splitting in zero-net-magnetization magnets, paving the way for the advancement and construction of low-power spintronic device .

\end{abstract}
\maketitle
\textcolor[rgb]{0.00,0.00,1.00}{\textbf{Introduction.---}}
Magnetism is a cornerstone of modern technology, and among magnetic materials, ferromagnets have long commanded the spotlight, driving decades of intensive research and ubiquitous applications\cite{1a}. A paradigm shift, however, is emerging with zero-net-magnetization systems. These magnets-free from any stray fields-deliver superior spintronic performance: ultrahigh data densities, immunity to external perturbations, and femtosecond-scale writing speeds\cite{k1,k2}.
Symmetry dictates that collinear, zero-net-magnetization magnets fall into three principal categories: $PT$-antiferromagnets (the joint symmetry ($PT$) of space inversion symmetry ($P$) and time-reversal symmetry ($T$)), alternagnets and fully-compensated ferrimagnets\cite{k4,k5,zg2,f4}.

 \begin{figure}[t]
    \centering
    \includegraphics[width=0.30\textwidth]{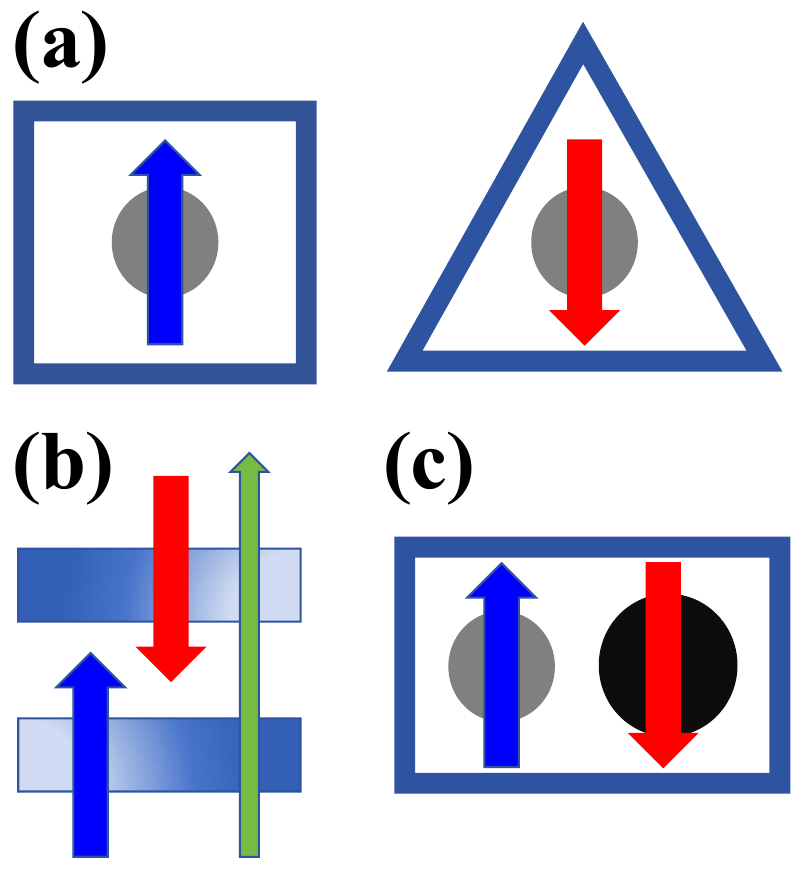}
    \caption{(Color online)(a):the magnetic atoms with opposite spin polarization have different environment, producing  fully-compensated ferrimagnetism. (b): the layer-dependent opposite spin polarization in $PT$-antiferromagnets and altermagnets can give rise to fully compensated ferrimagnetism under an applied electric field-whether it is an external field, the built-in field arising from a Janus structure, or the internal field of ferroelectric polarization. (c): in $PT$-antiferromagnets or altermagnets, the fully-compensated ferrimagnetism can be induced via  isovalent alloying.  In (a, b, c), the blue and red arrows represent spin-up and spin-down, respectively. In (b), the green arrow denotes the electric field. In (c), for example, the small gray sphere and the large black sphere represent 3$d$ and 4$d$ elements from the same chemical group, respectively.}\label{a0}
    \end{figure}

In $PT$-antiferromagnets, the $PT$ symmetry  enforces global spin degeneracy across the entire Brillouin zone (BZ), forbidding such hallmark phenomena as the magneto-optical response, anomalous Hall effect, and anomalous valley Hall effect\cite{zg2,zg1}. Alternagnets, by contrast, retain the real-space appearance of a collinear antiferromagnet-every spin is microscopically compensated-yet in momentum space they inherit and extend the essence of ferromagnetism. Without invoking relativistic effects, they exhibit robust spin splitting whose symmetry is classified by $d$-, $g$-, or $i$-wave representations\cite{k4,k5}. Consequently, alternagnets have been shown to host a suite of effects once regarded as the exclusive domain of ferromagnets: non-relativistic lifting of Kramers degeneracy, anomalous Hall and Nernst responses, spin-polarized charge currents, and the magneto-optical Kerr effect\cite{zg1}.

 \begin{figure}[t]
    \centering
    \includegraphics[width=0.40\textwidth]{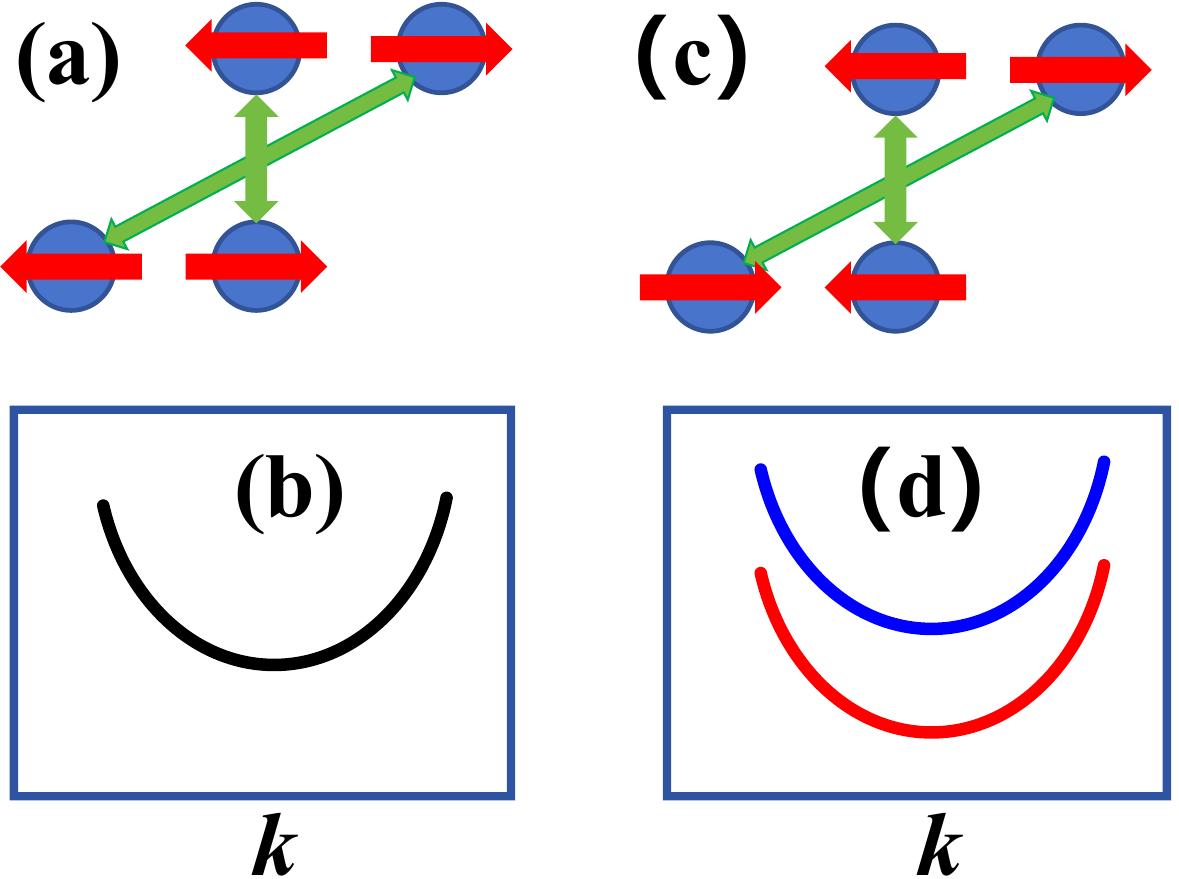}
    \caption{(Color online)Ignoring the spin ordering, both (a) and (c) possess lattice $P$ symmetry. When considering the spin ordering, (a) has   [$C_2$$\parallel$$P$]  symmetry, while (c) does not have   [$C_2$$\parallel$$P$]  symmetry and also lacks   [$C_2$$\parallel$$M$]  and   [$C_2$$\parallel$$C$]  symmetry. The (b) and (d) are the energy band structures corresponding to (a) and (c), respectively. In (a) and (c), the red arrows represent the spin direction, while the green arrows indicate that two atoms are connected by $P$ symmetry. In (b) and (d), the black lines represent spin degeneracy, while the blue and red lines represent spin-up and spin-down states, respectively.}\label{a}
\end{figure}

Fully-compensated ferrimagnets constitute a distinct branch of ferrimagnetism whose macroscopic moment vanishes\cite{f4,f1,f2,f3}.  Across the entire BZ, they exhibit a momentum-independent, $s$-wave spin splitting. Like alternagnets, fully-compensated ferrimagnets therefore give rise to phenomena: the anomalous Hall and Nernst effects, non-relativistic spin-polarized currents and the magneto-optical Kerr effect. Very recently, bilayer  $\mathrm{CrPS_4}$  can be electrically engineered into a  fully-compensated ferrimagnet: a perpendicular gate field toggles the conduction-band spin polarization on and off, offering an all-electric spin switch\cite{nn}.
Fully-compensated ferrimagnets can be obtained by breaking both the  [$C_2$$\parallel$$P$] symmetry of $PT$-antiferromagnets and the [$C_2$$\parallel$$C$]  or   [$C_2$$\parallel$$M$] symmetry of  alternagnets, where $C_2$ is the two-fold rotation perpendicular to the spin axis in spin space, and $C$/$M$  means rotation/mirror  symmetry in lattice space\cite{zg2,f4,qq3}

Previous strategies have always relied on engineering the lattice so that spin-opposite magnetic atoms reside in inequivalent environments\cite{gsd}-i.e., with different surrounding atomic arrangements-to realize fully-compensated ferrimagnets (See \autoref{a0} (a)). For example, the layer-dependent opposite spin polarization exists  in $PT$-antiferromagnets and altermagnets, which can  transform into fully-compensated ferrimagnets under an applied electric field-whether it is an external field, the built-in field arising from a Janus structure, or the internal field of ferroelectric polarization (See \autoref{a0} (b)) \cite{zg2,f4,qq3}. For instance, fully-compensated ferrimagnetism can also be achieved via isolectronic alloying (See \autoref{a0} (c))\cite{zg2,f4,f1,f2,f3,qq3}.
Indeed, in both $PT$-antiferromagnets and altermagnets, fully-compensated ferrimagnetism can also be realized by perturbing the environment around one spin-polarized atom through doping or vacancies\cite{gsd,gsd1}. A natural question is whether alternative strategies exist to realize fully-compensated ferrimagnetism. Here, we propose an alternative route to fully-compensated ferrimagnetism: instead of tailoring the lattice, we engineer the spin ordering.

\begin{figure}[t]
    \centering
    \includegraphics[width=0.40\textwidth]{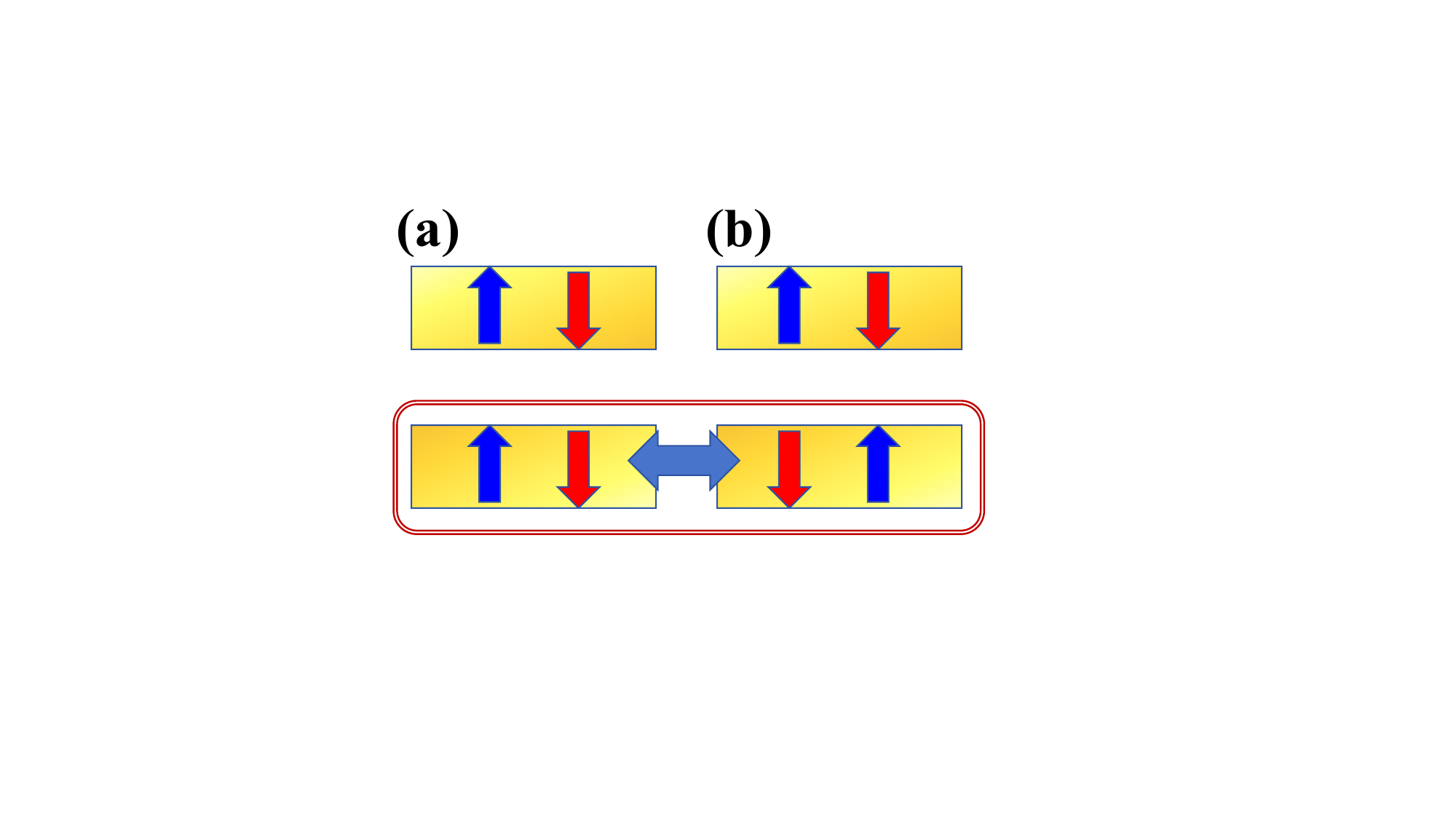}
    \caption{(Color online)By bilayer stacking engineering,  (a) has   [$C_2$$\parallel$$P$]  symmetry, while (b) does not have   [$C_2$$\parallel$$P$]  symmetry and also lacks   [$C_2$$\parallel$$M$]  and   [$C_2$$\parallel$$C$]  symmetry. The transition between (a) and (b) can be realized by controlling the  $\mathrm{N\acute{e}el}$ vector of the lower layer to flip by 180$^{\circ}$.   The blue and red arrows represent spin-up and spin-down, respectively.}\label{a1}
\end{figure}
\begin{figure*}[t]
    \centering
    \includegraphics[width=0.80\textwidth]{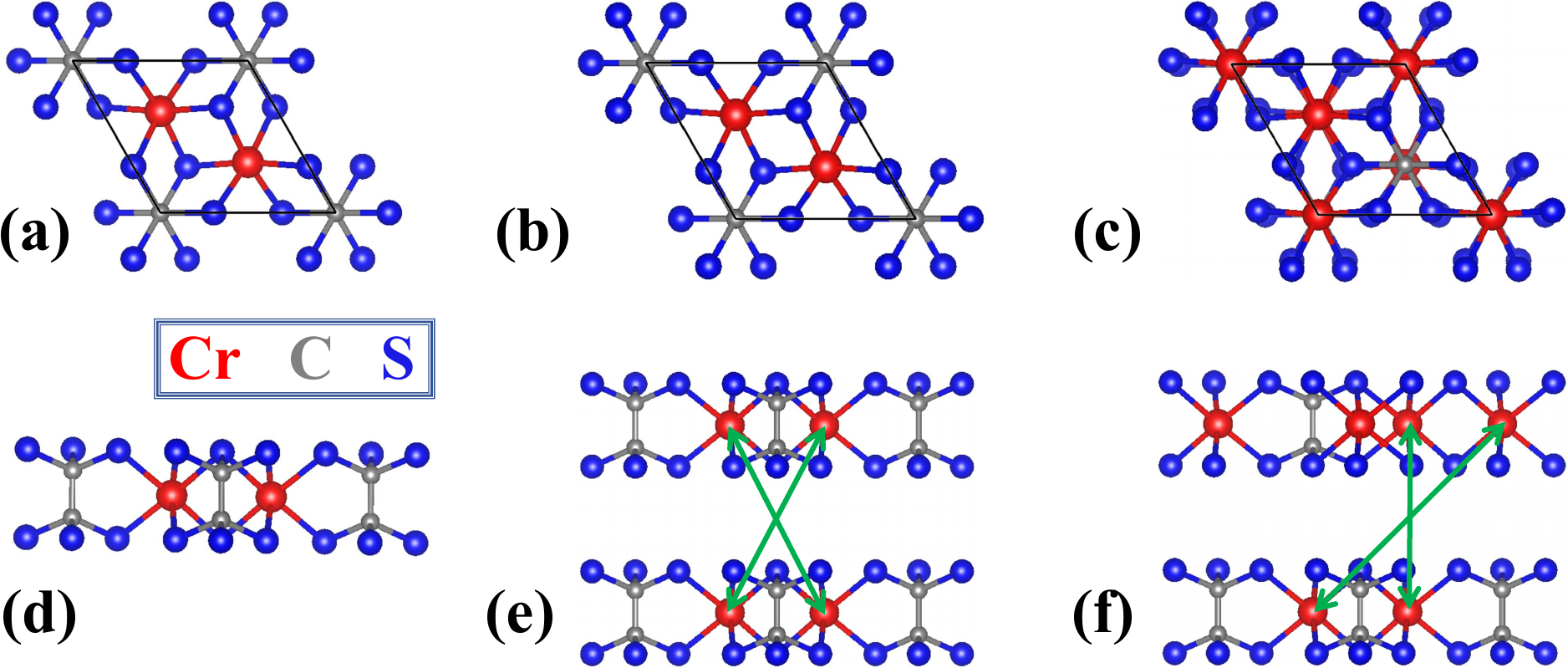}
    \caption{(Color online)The top (a, b, c) and side (d, e, f) views of the crystal structures of monolayer $\mathrm{Cr_2C_2S_6}$ (a, d), bilayer A-stacked $\mathrm{Cr_2C_2S_6}$ (b, e) and bilayer B-stacked $\mathrm{Cr_2C_2S_6}$ (c, f). In (e) and (f),  the green arrows indicate that two Cr atoms are connected by $P$ symmetry.}\label{b}
\end{figure*}

\textcolor[rgb]{0.00,0.00,1.00}{\textbf{Approach.---}}
To achieve spin ordering-induced fully-compensated ferrimagnetism, the primitive unit cell should generally contain at least four magnetic atoms, as shown in \autoref{a} (a) and (c).  When ignoring the spin ordering, both \autoref{a} (a) and (c) possess lattice $P$ symmetry.   When considering the spin ordering, \autoref{a} (a)  has   [$C_2$$\parallel$$P$]  symmetry, which leads to  spin degenerate due to  [$C_2$$\parallel$$P$][$C_2$$\parallel$$T$] symmetry ($E_{\uparrow}(k)$=[$C_2$$\parallel$$P$][$C_2$$\parallel$$T$]$E_{\uparrow}(k)$= [$C_2$$\parallel$$P$]$E_{\uparrow}(-k)$=$E_{\downarrow}(k)$), as shown in  \autoref{a} (b). In fact, \autoref{a} (a) shows  $PT$-antiferromagnetism.
When considering the spin ordering,  \autoref{a} (c) does not possess   [$C_2$$\parallel$$P$]  symmetry.
This broken   [$C_2$$\parallel$$P$]  symmetry will induce spin splitting. However, if [$C_2$$\parallel$$M/C$]   symmetry is present, it will lead to alternemagnetism ($E_{\uparrow}(k)$=[$C_2$$\parallel$$M/C$]$E_{\uparrow}(k)$=$E_{\downarrow}(M/Ck)$)\cite{k4,k5}. If these two symmetries are also excluded, it will result in fully-compensated ferrimagnetism with  global spin splitting (see \autoref{a} (d))\cite{f4}.
From \autoref{a} (a) to (c), simply changing the spin ordering without altering the atomic arrangement can induce fully-compensated ferrimagnetism,  achieving global spin splitting.

It may be difficult to directly find the material we propose. Here, we achieve our proposal through bilayer stacking engineering, which   has already served as a versatile platform for realizing diverse electronic states\cite{o1,o2,o3,o4}.
For example, we first search  for monolayer that contain two magnetic atoms with antiferromagnetic (AFM) coupling, which  is used as the basic building unit. Subsequently, by employing operations such as reflection, rotation, and translation on the basic building unit, we construct a bilayer that fulfills the lattice $P$ symmetry.
The spin-degenerate and globally spin-splitting states, that is, $PT$-antiferromagnetism and fully-compensated ferrimagnetism, can be interconverted by manipulating the $\mathrm{N\acute{e}el}$ vector of the lower layer, specifically by flipping the $\mathrm{N\acute{e}el}$ vector by 180$^{\circ}$ (see \autoref{a1}).
Experimentally, the switching of $\mathrm{N\acute{e}el}$ vector can be realized by spin-orbit torques, and  the  reorientation of $\mathrm{N\acute{e}el}$ vectors with  180$^{\circ}$ switching has been realized experimentally\cite{zg8}.
\begin{figure}[t]
    \centering
    \includegraphics[width=0.45\textwidth]{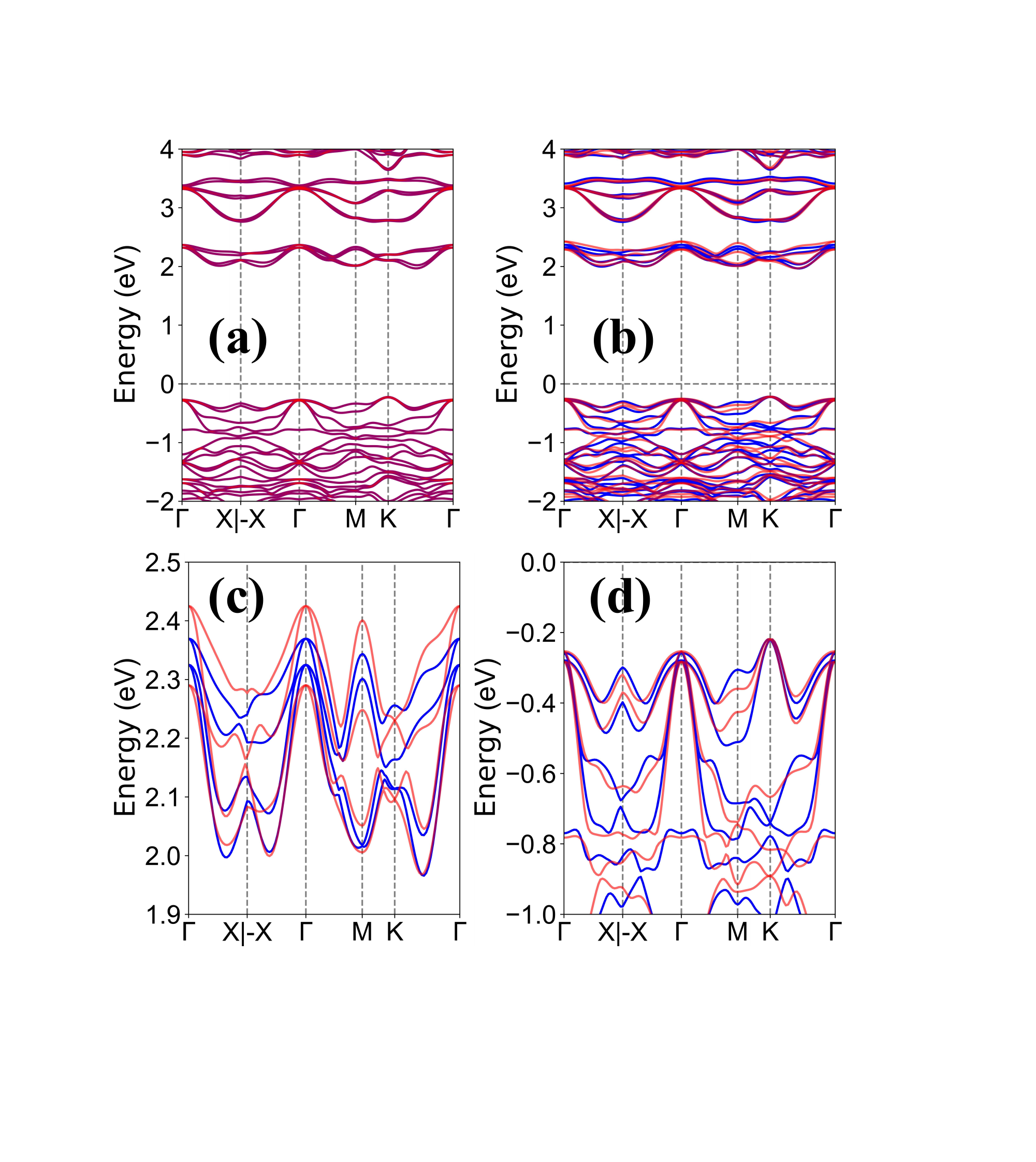}
     \caption{(Color online) The energy band structures of bilayer B-stacked $\mathrm{Cr_2C_2S_6}$ with $PT$ symmetry (a) and without  $PT$ symmetry (b). The  (c) and (d) are the enlargements of the conduction and valence bands near the Fermi level for (b). The spin-up and spin-down channels are depicted in blue and red, and the  purple color means spin degeneracy.}\label{c}
\end{figure}

\textcolor[rgb]{0.00,0.00,1.00}{\textbf{Computational detail.---}}
The spin-polarized first-principles calculations are  performed within density functional theory (DFT)\cite{1}using the Vienna Ab Initio Simulation Package (VASP)\cite{pv1,pv2,pv3}.  We employ the  Perdew-Burke-Ernzerhof generalized gradient approximation (PBE-GGA)\cite{pbe}  as the exchange-correlation functional. The calculations are carried out with the kinetic energy cutoff  of 500 eV,  total energy  convergence criterion of  $10^{-8}$ eV, and  force convergence criterion of 0.001 $\mathrm{eV.{\AA}^{-1}}$. To account for the localized nature of Cr-3$d$ orbitals,  we use a Hubbard correction $U_{eff}$  within  the rotationally invariant approach proposed by Dudarev et al\cite{du}. The $U_{eff}$=3.0 eV\cite{qq3,f7-1} and 3.55 eV\cite{o5-1,o5} are  used for $\mathrm{Cr_2C_2S_6}$ and $\mathrm{Cr_2SO}$, respectively. A vacuum layer exceeding 16 $\mathrm{{\AA}}$ along the $z$-direction is employed to eliminate spurious interactions between periodic images. The BZ is sampled with a 15$\times$15$\times$1 Monkhorst-Pack $k$-point meshe for both structural relaxation and electronic structure calculations.

\begin{figure*}[t]
    \centering
    \includegraphics[width=0.95\textwidth]{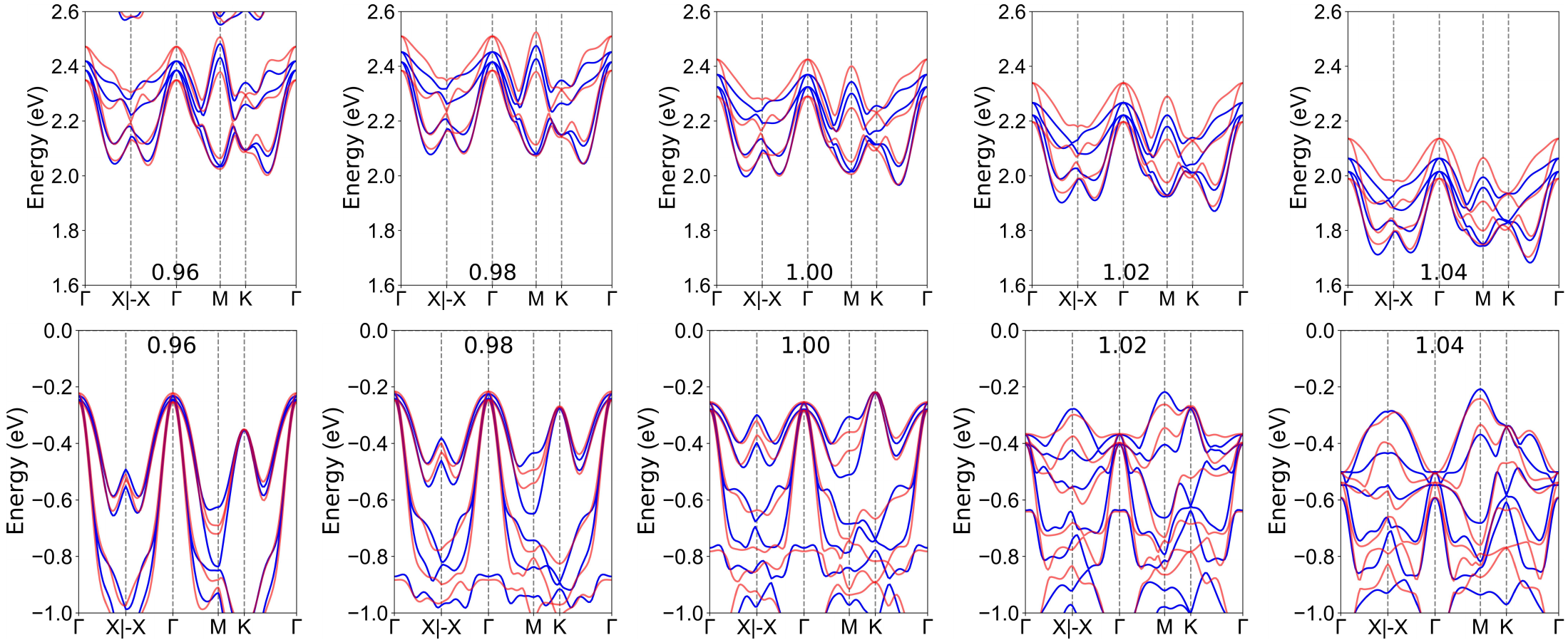}
     \caption{(Color online) The enlargements of the conduction (top) and valence (bottom) bands near the Fermi level for bilayer B-stacked $\mathrm{Cr_2C_2S_6}$  without  $PT$ symmetry at representative $a/a_0$ (0.96, 0.98, 1.00, 1.02 and 1.04). The spin-up and spin-down channels are depicted in blue and red. }\label{d}
\end{figure*}

\begin{figure}[t]
    \centering
    \includegraphics[width=0.45\textwidth]{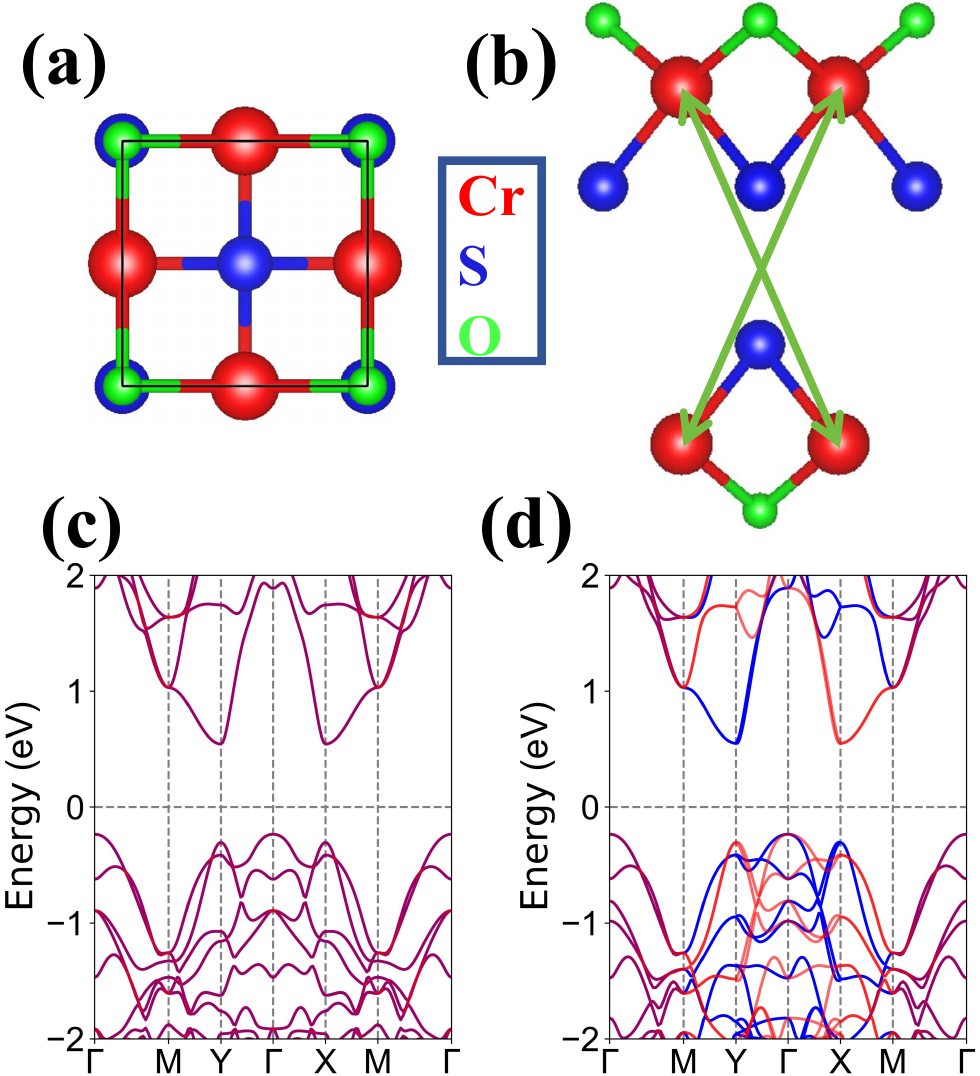}
     \caption{(Color online)The top (a) and side (b) views of the crystal structures of  bilayer B-stacked $\mathrm{Cr_2SO}$ . In (b),  the green arrows indicate that two Cr atoms are connected by $P$ symmetry. The energy band structures of bilayer B-stacked $\mathrm{Cr_2SO}$ with $PT$ symmetry (c) and without  $PT$ symmetry but  [$C_2$$\parallel$$M$] symmetry (d). In (c) and (d), the spin-up and spin-down channels are depicted in blue and red, and the  purple color means spin degeneracy. }\label{e}
\end{figure}

\textcolor[rgb]{0.00,0.00,1.00}{\textbf{Material realization.---}}
Here, the monolayer $\mathrm{Cr_2C_2S_6}$ is used  the basic building unit, which has been proved to be dynamically, mechanically and thermally stable\cite{qq3}.
As shown in \autoref{b} (a) and (d),  there are two Cr atoms in the primitive unit cell, which  are surrounded by six S
atoms, and the  two $\mathrm{CrS_3}$ moieties are connected by two C atoms. The $\mathrm{Cr_2C_2S_6}$ crystallizes in the  $P\bar{3}1m$ space group (No.162) with  $P$ and $M$ symmetries, which  transform one Cr sublattice into the other.
 The magnetic  ground state of $\mathrm{Cr_2C_2S_6}$ among ferromagnetic (FM),  AFM-N$\acute{e}$el  (AFM1),  AFM-stripy (AFM2) and  AFM-zigzag (AFM3) magnetic configurations is AFM1 ordering (See FIG.S1\cite{bc}), and the energy differences between  FM/AFM2/AFM3 and AFM1 orderings  are  266 meV,  167 meV and 114 meV, respectively. Within AFM1 ordering, the  optimized  lattice constants are  $a$=$b$=5.636 $\mathrm{{\AA}}$  for monolayer $\mathrm{Cr_2C_2S_6}$.

Then, we construct the bilayer structure, with $\mathrm{Cr_2C_2S_6}$ as the lower layer, and the upper layer can be obtained through a horizontal mirror operation,  which is named as bilayer A-stacked $\mathrm{Cr_2C_2S_6}$,  as shown in  \autoref{b} (b) and (e).  The bilayer A-stacking  also   crystallizes in the  $P\bar{3}1m$ space group (No.162), possessing lattice $P$ symmetry. However, after the  [$C_2$$\parallel$$P$] is broken by  tuning the spin ordering, it still possesses [$C_2$$\parallel$$M$], which will result in alternemagnetism, a subject to be discussed later. In addition, the energy of this type of bilayer stacking is generally not a minimum. By translating the top sublayer by   ($\vec{a}$/3+2$\vec{b}$/3),  the bilayer B-stacked $\mathrm{Cr_2C_2S_6}$ can be obtained, as shown in  \autoref{b} (c) and (f). The B-stacking indeed has an energy 22 meV lower than the A-stacking. Moreover, the B-stacking still preserves the lattice $P$ symmetry with  $P\bar{3}$ space group (No.147).

Because the AFM1 ordering of monolayer $\mathrm{Cr_2C_2S_6}$ is at least 114 meV lower in energy than any other magnetic configuration, and the interlayer coupling is extremely weak, we keep each layer $\mathrm{Cr_2C_2S_6}$ locked in the AFM1 ordering and only vary the interlayer magnetic arrangements.
For  B-stacking, the magnetic ordering with $PT$ symmetry is 1.04 meV lower in energy than that without $PT$ symmetry. Nevertheless, given the tiny energy difference between them, the two magnetic states can be interconverted simply by tuning the $\mathrm{N\acute{e}el}$ vector  of the lower layer $\mathrm{Cr_2C_2S_6}$. For B-stacking, once the spin order breaks   [$C_2$$\parallel$$P$] symmetry, neither   [$C_2$$\parallel$$M$] nor   [$C_2$$\parallel$$C$] symmetries survive, so the bilayer B-stacked $\mathrm{Cr_2C_2S_6}$ becomes a fully-compensated ferrimagnet.

Next, we corroborate our analysis by examining the energy band structures. Since monolayer  $\mathrm{Cr_2C_2S_6}$ exhibits altermagnetism with  $i$-wave spin-splitting symmetry under an out-of-plane electric field\cite{qq3}, we have added the $\Gamma$-X$\mid$X'-$\Gamma$ path\cite{qq1} for the related  energy band structures, which is shown in FIG.S2\cite{bc}.
The energy band structures of bilayer B-stacked $\mathrm{Cr_2C_2S_6}$ with $PT$ symmetry  and without  $PT$ symmetry, along with the enlargements of the conduction and valence bands near the Fermi level for the non-$PT$-symmetric case, are plotted in \autoref{c}.

For the $PT$-symmetric case, the band structures clearly exhibit spin degeneracy.
For the non-$PT$-symmetric case, the bands show obvious spin splitting, and  the band structures exhibit no symmetric connectivity along the high-symmetry paths $\Gamma$-X and X'-$\Gamma$, which is different from the case of altermagnetism. Together with the fact that the total magnetic moment of the bilayer $\mathrm{Cr_2C_2S_6}$  is 0 $\mu_B$, this confirms that bilayer B-stacked $\mathrm{Cr_2C_2S_6}$ is indeed a fully-compensated ferrimagnet.  When  fully-compensated ferrimagnetism is realized via an electric field, Janus engineering or alloying, the absolute values of the magnetic moments on the magnetic atoms are unequal\cite{zg2,f4}.  Nevertheless, the spin ordering-induced fully-compensated ferrimagnetism yields identical absolute values of  magnetic moments (3.006 $\mu_B$) on all Cr atoms in bilayer $\mathrm{Cr_2C_2S_6}$.

Strain can effectively  tune the electronic structure and  magnetic properties of two-dimensional (2D) materials.
Here,  the ratio $a/a_0$ (The $a$ and $a_0$ ​represent the strained and unstrained lattice constants, respectively)  is employed to simulate biaxial strain, where $a/a_0$$<$1 ($a/a_0$$>$1) corresponds to compressive (tensile) strain. Within considered strain range ($a/a_0$=0.96$\backsim$1.04),  the magnetic ordering with $PT$ symmetry is always  lower in energy than that without $PT$ symmetry (See FIG.S3\cite{bc}).
The enlargements of the conduction and valence bands near the Fermi level for bilayer B-stacked $\mathrm{Cr_2C_2S_6}$  without  $PT$ symmetry at representative $a/a_0$ are shown in \autoref{d}. A pivotal strain-induced effect is the controllable relocation of the valence band maximum (VBM): compressive strain drives the VBM to the $\Gamma$ point, whereas tensile strain shifts it to the M point. The pronounced spin splitting at the M point of the valence band makes tensile strain more favorable for spintronic applications (See the band structure at 1.02  strain).

\textcolor[rgb]{0.00,0.00,1.00}{\textbf{Discussion and conclusion.---}}
In addition to spin ordering giving rise to fully-compensated ferrimagnetism, it can also induce altermagnetism.
 When \autoref{a} (c) does not possess   [$C_2$$\parallel$$P$]  symmetry,  but  [$C_2$$\parallel$$M$]  or  [$C_2$$\parallel$$C$] symmetry is present, it will lead to alternemagnetism. Likewise, we employ bilayer stacking engineering to corroborate our proposal.
 Using monolayer $\mathrm{Cr_2SO}$\cite{o5}as the basic building unit, we construct S-terminal bilayer and O-terminal bilayer with  lattice $P$ symmetry  as A stacking  via mirror symmetry operation, which crystallizes in the $P4/mmm$ space group (No.123).
 The energy of  O-terminal bilayer is  15 meV lower than that of  S-terminal bilayer. We will focus  on  O-terminal bilayer,  and the crystal structures of  O-terminal bilayer with A stacking along with monolayer  $\mathrm{Cr_2SO}$ are plotted in FIG.S4\cite{bc}.

To obtain lower-energy stacking configuration, based on  A stacking, the B stacking  (See \autoref{e} (a) and (b)) is obtained by translating the top sublayer by  ($\vec{a}$+$\vec{b}$)/2 along the diagonal direction. The B stacking crystallizes  in $P4/nmm$ (No.129), and it has also  lattice $P$ symmetry. It is found that the energy of B stacking  is approximately 114 meV lower than that of A stacking.
For  B-stacking, the magnetic ordering with $PT$ symmetry is 1.13 meV higher in energy than that without $PT$ symmetry.  It is noteworthy that a magnetic ordering without $PT$ symmetry can  host altermagnetism when  [$C_2$$\parallel$$P$]  symmetry is absent but  [$C_2$$\parallel$$M$]  symmetry is preserved.
The energy band structures of bilayer B-stacked $\mathrm{Cr_2SO}$ with $PT$ symmetry and without  $PT$ symmetry are plotted in \autoref{e} (c) and (d).
It is evident that tuning the spin ordering enables a transition between $PT$-antiferromagnetism and altermagnetism.
Experimentally, the transition between the two magnetic states can also be achieved by flipping the $\mathrm{N\acute{e}el}$ vector of the lower layer $\mathrm{Cr_2SO}$. In fact, bilayer A-stacked $\mathrm{Cr_2C_2S_6}$ can likewise undergo a transition between $PT$-antiferromagnetism and altermagnetism via spin ordering tuning (See FIG.S5\cite{bc}), yet its energy does not correspond to a minimum. Similarly, spin symmetry-induced altermagnetism has also been achieved in  $\mathrm{Mn_2P_2Te_6}$ bilayer\cite{o6}.

In summary,  we present  an alternative strategy for inducing fully-compensated ferrimagnetism by modulating
spin ordering, which is  further validated by first-principles calculations in bilayer  $\mathrm{Cr_2C_2S_6}$.
The two magnetic states with and without $PT$ symmetry can be interconverted simply by tuning the $\mathrm{N\acute{e}el}$ vector  of the lower or upper layer $\mathrm{Cr_2C_2S_6}$. Spin ordering-driven fully-compensated ferrimagnetism equalizes the absolute magnetic moments on every Cr atom in bilayer $\mathrm{Cr_2C_2S_6}$. This differs fundamentally from the conventional structure-modulation route, where the absolute magnetic moments of the magnetic atoms in a fully-compensated ferrimagnet are unequal. These findings establish a clear path toward exploring fully-compensated ferrimagnetism by tuning spin ordering.

\begin{acknowledgments}
This work is supported by Natural Science Basis Research Plan in Shaanxi Province of China  (2025JC-YBMS-008). We are grateful to Shanxi Supercomputing Center of China, and the calculations were performed on TianHe-2.
\end{acknowledgments}

\end{document}